\documentclass{aa}
\usepackage{graphicx}

\def\gtrsim{\mathrel{\hbox{\rlap{\hbox{\lower4pt\hbox{$\sim$}}}\hbox{$>$}}}}
\def\ltsim{\mathrel{\hbox{\rlap{\hbox{\lower4pt\hbox{$\sim$}}}\hbox{$<$}}}}

\usepackage{color}

\begin{document}

\title{14 Ceti: A probable Ap-star-descendant entering the Hertzsprung gap. \thanks{Based on data obtained using the T\'elescope Bernard Lyot at Observatoire du 
Pic du Midi, CNRS/INSU and Universit\'e de Toulouse, France.}}

\author{M. Auri\`ere\inst{1,2}, R. Konstantinova-Antova\inst{3,1},  P. Petit\inst{1,2}, C. Charbonnel\inst{4,2}, S. Van Eck\inst{5}, J.-F. Donati\inst{1,2}, F. Ligni\`eres\inst{1,2}, and T. Roudier\inst{1,2}  
}
\offprints{michel.auriere@ast.obs-mip.fr}
\institute{Universit\'e de Toulouse, UPS-OMP, Institut de Recherche en Astrophysique et Plan\'etologie, Toulouse, France\\
\and
CNRS, Institut de Recherche en Astrophysique et Plan\'etologie, 14 Avenue Edouard Belin, 31400 Toulouse, France\\
\email{[michel.auriere;ppetit;jean-francois.donati;francois.lignieres;thierry.roudier]@irap.omp.eu}
\and
Institute of Astronomy and NAO, Bulgarian Academy of Sciences, 72 Tsarigradsko shose, 1784 Sofia, Bulgaria\\
\email{renada@astro.bas.bg}
 \and
Geneva Observatory, University of Geneva, 51 Chemin des Maillettes, 1290 Versoix, Switzerland\\
\email{corinne.charbonnel@unige.ch}
 \and
Institut d'Astrophysique, Universit\'e libre de Bruxelles, Campus Plaine - C.P. 226, 1050 Bruxelles, Belgium\\
\email{svaneck@ulb.ac.be}
 }

 \date{Received ??; accepted ??}

\abstract
 {14 Ceti is a subgiant star of F spectral class that displays variations in the $S$-index of its Ca~{\sc ii} H \& K lines and an X-ray emission that is stronger than the mean observed for its spectral class, which may be due to some magnetic activity.}
{We attempt to Zeeman-detect and study the magnetic field of 14 Ceti and to infer its origin.}
{We used the spectropolarimeter Narval at the Telescope Bernard Lyot, Pic du Midi Observatory, 
and the least squares deconvolution  method to create high signal-to-noise ratio Stokes $V$ profiles. We derived the surface-averaged longitudinal magnetic field $B_{l}$. We also measured the $S$-index, and the radial velocity for each observation.}
{14 Ceti is Zeeman-detected for the 30 observed dates spanning from August 2007 to January 2012. The average longitudinal magnetic field does not reverse its sign, reaches about -35 G,  and shows some month-long-timescale variations in our 2008 and 2011-2012 observations. The $S$-index follows the same long-term trend as $B_l$. 14 Ceti is confirmed as a single star without H-K emission cores. The strength of the observed surface magnetic field of 14 Ceti is one order of magnitude  greater than the observed one for late F main-sequence stars, and is comparable to the values measured in the active late F pre-main-sequence star HR 1817. On the other hand, taking into account the post-main-sequence evolution of an Ap star, an oblique rotator model can explain the strength of the magnetic field of 14 Ceti. The variations with a timescale of months observed for both the $B_l$ and $S$-index could be due to the rotation.}
{The most probable scenario to explain our observations appears to be that 14 Ceti is the descendant of a cool Ap star  .}

   \authorrunning {M. Auri\`ere et al.}
   \titlerunning {The magnetic field of 14 Ceti}

\maketitle

\section{Introduction}

14 Ceti (HD 3229, HR 143) has been classified  as F5V (Gray 1989) and either F5IV or F5IV-V (Hoffleit \& Warren 1991). 
It was included as a ``single star with no H-K emission'' in a sample devoted to the ``attempt to connect the rotations of main-sequence stars with their chromospheric properties'' (Wilson 1966). It was observed over 25 years (1966-1991) as part of the HK Mount Wilson survey (Duncan et al. 1991, Baliunas et al. 1995). 
Baliunas et al. (1995) inferred a possible activity cycle of about five years, but with a poor confidence level. Noyes et al. (1984) predicted a rotational period of 2.4 days for 14 Ceti from the CaII H \& K properties.
14 Ceti appears in the ``ROSAT all-sky survey catalogue of optically bright main-sequence stars and subgiant stars'' (H\"unsch et al. 1998). Its X-ray luminosity of 0.33 $10^{30} \mathrm{\,erg\,s^{-1}}$  is rather strong for an F5-type star (Schmitt et al. 1985), but is not  unique in the ROSAT catalogue (H\"unsch et al. 1998). Its [$L_{\rm X}/L_{\rm bol}$] as presented by Bruevich et al. (2001) is similarly also high.

 Since the properties of its chromosphere and corona are indicative of magnetic activity, we included 14 Ceti in a magnetic (Zeeman) survey of evolved single intermediate-mass stars with Narval and detected its surface magnetic field (Auri\`ere et al. 2009a). 14 Ceti then became the only star between F0 and F7 spectral types to be Zeeman-detected. It could then be either the hottest known star hosting a dynamo-driven surface magnetic field, or another example of a descendant of an Ap-star, such as EK Eri (e.g. Stepie\' n et al. 1993, Auri\`ere et al. 2011). We present here the Zeeman investigation that we have conducted
to infer the origin of the magnetic activity of 14 Ceti, and discuss its possible evolutionary and magnetic status as inferred from stellar evolution models.

\section{Observations with Narval}

\subsection{Observations}

 We used Narval at the TBL (2m telescope Bernard Lyot at Observatoire du Pic du Midi, France), which is a copy of the new generation spectropolarimeter
ESPaDOnS (Donati et al. 2006a). Narval consists of a Cassegrain polarimetric module connected by optical fibres to an echelle spectrometer. In polarimetric mode, the instrument simultaneously acquires two orthogonally polarized spectra covering the spectral range from 370 nm to 1000 nm in a single exposure, with a resolving power of about 65000. 

A standard circular-polarization observation consists of a series of four sub-exposures between which the 
 half-wave retarders (Fresnel rhombs) are rotated to exchange the paths of the orthogonally polarized beams within the whole instrument (and therefore the positions of the two spectra on the CCD), thereby reducing the spurious polarization signatures. The extraction of the spectra, including wavelength calibration, correction to the heliocentric frame, and continuum normalization, was performed using Libre-ESpRIT (Donati et al. 1997), a dedicated and automatic reduction package installed at TBL. The extracted spectra are produced in ASCII format, and consist of the normalised Stokes $I$ ($I/I_{\rm c}$) and Stokes $V$ ($V/I_{\rm c}$) parameters as a function of wavelength , along with their associated Stokes $V$ uncertainty $\sigma_V$ (where $I_{\rm c}$ represents the continuum intensity). We also include in the output "diagnostic null" spectra $N$, which are in principle featureless, and therefore serve to diagnose the presence of spurious contributions to the Stokes $V$ spectrum.

14 Ceti was observed on 7 nights between August 2007 and December 2008, then 23 nights during a survey in the 2011-2012 season. These observations are listed in Table 1. The exposure time for each observation was 40 min and the maximum signal-to-noise ratio (S/N) in Stokes $I$ per 2.6 $\mathrm{\,km\,s^{-1}}$ spectral bin was between 490 and 1000.

To complete the Zeeman analysis, least squares deconvolution 
(LSD, Donati et al. 1997) was applied to all observations. This is a multi-line technique, similar to cross-correlation,  which assumes that all spectral lines have the same profile, scaled by a certain factor. We used a digital mask with about 6,500 photospheric lines, calculated for an effective 
temperature of  6500 K, and $\log g$ = 4, using the Kurucz models (1993). In the present case, this method enabled us to derive
Stokes V profiles with a S/N  that was higher by a factor of about 30 than those for single lines.  We 
then computed the longitudinal magnetic field $B_{l}$ in G, using the 
first-order moment method (Rees \& Semel 1979, Donati et al. 1997). Kochukhov et al. (2010) investigated the LSD method, showing that it provides estimates of the longitudinal magnetic field that are accurate to within a few percent for fields weaker than 1 kG, as which are considered here.

The activity of the star during the same nights was monitored by computing the $S$-index (defined from the Mount Wilson survey, Duncan et al. 1991) for the chromospheric Ca~{\sc ii} H \& K line cores on our spectra. Our procedure was calibrated using the main-sequence solar-type stars of Wright et al. (2004). 

We measured the radial velocity ($RV$) of 14 Ceti on the LSD Stokes $I$ profiles using a Gaussian fit. The long-term stability of Narval is about 30 $\mathrm{\,m\,s^{-1}}$ (e.g. Moutou et al. 2007, Auri\`ere et al. 2009b) and the absolute accuracy of individual measurements relative to the local standard of rest is about 1 $\mathrm{\,km\,s^{-1}}$. 

\begin{figure}
\centering
\includegraphics[width=8 cm,angle=0] {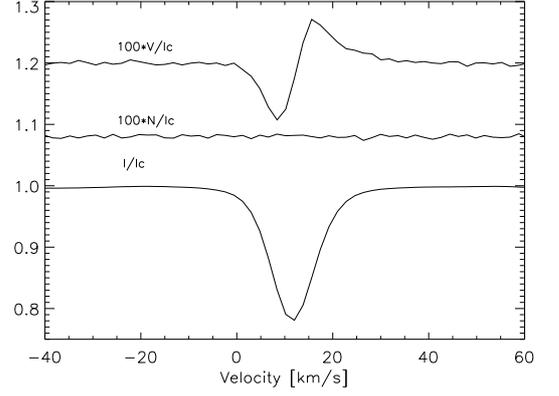} 
\caption{Mean LSD profiles of 14 Ceti as observed with Narval on 25 Sep. 2008. Stokes $V$ (upper), $N$ nul polarization (middle), and Stokes $I$ (lower) profiles are presented. For display purposes, the profiles are shifted vertically, and the Stokes $V$ as well as the $N$ profiles are expanded by a factor of 100.}
\end{figure}

Table 1 provides the date, HJD, the CaII H \& K  $S$- index, $B_l$ value and its uncertaintiy in G, and $RV$.

\begin{table}
\caption{Observations of 14 Ceti (for details, see Sect. 2.1) }           
\centering                         
\begin{tabular}{c c c c c c}     
\hline\hline               
Date        & HJD      &$S$-ind.& $B_l$& $\sigma$ & $RV$       \\
            &2450000+  &          & G    &  G       & km s$^{-1}$ \\
\hline                        
18 Aug. 2007&4331.61  &0.224    &-07.82&1.30      &11.685 \\
14 Sep. 2008&4724.48  &0.231    &-31.88&1.57      &11.695 \\
16 Sep. 2008&4726.45  &0.232    &-31.31&1.27      &11.670 \\
19 Sep. 2008&4729.53  &0.232    &-31.22&1.92      &11.680 \\
20 Sep. 2008&4730.41  &0.234    &-32.63&2.15      &11.640 \\
25 Sep. 2008&4735.46  &0.231    &-28.24&1.23      &11.655 \\
21 Dec. 2008&4822.28  &0.229    &-17.06&1.21      &11.545 \\
22 Jul. 2011&5765.63  &0.234    &-25.20&1.20      &11.576 \\
27 Aug. 2011&5801.59  &0.235    &-25.81&1.69      &11.690 \\
24 Sep. 2011&5829.48  &0.240    &-29.36&1.65      &11.688 \\
27 Sep. 2011&5832.53  &0.240    &-31.83&1.34      &11.660 \\
01 Oct. 2011&5836.46  &0.240    &-30.25&1.33      &11.709 \\
04 Oct. 2011&5839.55  &0.239    &-31.40&1.12      &11.658 \\
10 Oct. 2011&5845.42  &0.239    &-34.38&1.72      &11.677 \\
12 Oct. 2011&5847.38  &0.239    &-32.33&1.52      &11.668 \\
14 Oct. 2011&5849.49  &0.238    &-33.83&1.68      &11.648 \\
16 Oct. 2011&5851.39  &0.238    &-29.13&1.08      &11.625 \\
25 Oct. 2011&5860.44  &0.241    &-32.16&1.61      &11.683 \\
30 Oct. 2011&5865.41  &0.240    &-29.61&1.72      &11.684 \\
08 Nov. 2011&5874.36  &0.240    &-33.48&2.60      &11.696 \\
12 Nov. 2011&5878.47  &0.241    &-35.10&1.62      &11.710 \\
16 Nov. 2011&5882.44  &0.241    &-34.43&1.30      &11.709 \\
21 Nov. 2011&5887.40  &0.240    &-33.35&1.66      &11.720 \\
27 Nov. 2011&5893.28  &0.242    &-34.19&1.32      &11.714 \\
08 Dec. 2011&5904.29  &0.242    &-33.57&1.38      &11.636 \\
12 Dec. 2011&5908.34  &0.239    &-36.17&1.63      &11.672 \\
08 Jan. 2012&5935.25  &0.237    &-33.58&1.23      &11.570 \\
19 Jan. 2012&5946.27  &0.237    &-37.35&2.15      &11.658 \\
22 Jan. 2012&5949.25  &0.237    &-34.79&1.24      &11.687 \\
26 Jan  2012&5953.26  &0.237    &-32.48&1.24      &11.685 \\
\hline  
\end{tabular}
 
\end{table}

\subsection{Magnetic field, $S$-index, and $RV$ measurements}

The surface  magnetic field of 14 Ceti is Zeeman-detected on each observation described in Table 1. Figure 1 shows the LSD Stokes $V$, $N$ null polarization, and Stokes $I$ profiles obtained on 25 September 2008.  On each date, the Stokes $V$ profile does not display any peculiar details apart from the large-scale smooth behavior shown in Fig. 1. This can be explained by the small $v\sin i$ of 14 Cet, which is measured to be $5 \pm 1 \mathrm{\,km\,s^{-1}}$  (de Medeiros \& Mayor 1999). Table 1 shows that the average longitudinal magnetic field does not reverse its sign and varies from about -8 G (17 August 2007) to about -30 G in September 2008 and during September 2011-January 2012. 

Our observations confirm that 14 Ceti does not contain emission cores in the chromospheric Ca~{\sc ii} H \& K lines (e.g. Fig. 2). In reality, the core profiles are as deep as the ''wing-nibs'' described by Cowley et al. (2006) for main-sequence A-stars. The $S$-index values reported in Table 1, which vary between 0.224 and 0.242, are consistent with those of the HK Mount Wilson survey (Duncan et al. 1991), which vary between 0.200 and 0.242.  The $S$-index values for 14 Ceti are constant during the September 2008 ($S$-index $\approx$ 0.232) and September-December 2011 observations ($S$-index $\approx$ 0.240, see Fig. 3). The internal accuracy of the $S$-index is about 0.002, as illustrated by the fluctuations observed in these phases of apparent constant magnetic activity of 2008 and 2011.

The $RV$ varies within the range 11.640-11.700 $\mathrm{\,km\,s^{-1}}$ during the four years of observations. It is weaker by 100 $\mathrm{\,km\,s^{-1}}$ on 21 December 2008, 22 July 2011 and 19 January 2012. These differences can be due to different atmospheric conditions (Moutou et al. 2007). The stability of $RV$ confirms that 14 Ceti is a single star.

\subsection{Magnetic field and  $S$-index variations, and search for the rotational period}

Figure 3 plots the 30 $B_l$ (upper graph) and $S$-index (lower graph) measurements. This figure shows that the large timescale variations in $B_l$ and $S$-index are the same. On 18 August 2007, the weakest values were observed for both measurements. In September-December 2008, a general decrease in $B_l$ and $S$-index was measured. In the July - December 2011  observations, there is a general  increase, then a possible decrease during the January 2012 observations, in both the $B_l$ and $S$-index. If these variations were due to the rotational modulation, they would be indicative of a rather long rotational period, of a few months. 

\begin{figure}
\centering
\includegraphics[width=6 cm,angle=-90] {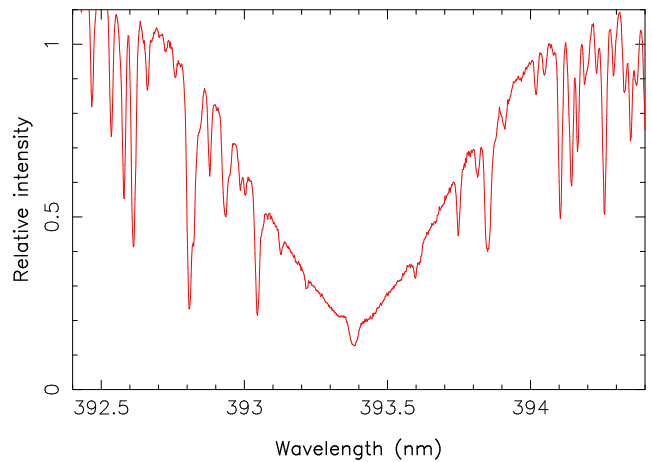} 
\caption{Spectrum of 14 Ceti on 25 Sep. 2008 showing the core of the chromospheric Ca~{\sc ii} K line.}
\end{figure}

Since Noyes et al. (1984)  predicted a short rotational period for 14 Ceti, we surveyed the magnetic field in September 2008 but did not observe  significant variations for four observations spanning along six days. We performed a new survey over the 23 nights of the 2011-2012 observational season of 14 Ceti, obtaining a more finely sampled and longer series. Table 1 and Fig. 4 (upper graph) show some marginally significant variations in $B_l$ during the September-January span of the 2011-2012 observations with a timescale of a few tens of days and an amplitude of about 3-4 G. Figure 4 (lower graph) shows that the $S$-index does not correlate with $B_l$ on this timescale. We then searched for periodic variations using the procedure described in  Petit et al. (2002) and the Zeeman Doppler imaging models of Donati et al. (2006b), in a way described in Auri\`ere et al. 2011. We could not determine any significant period in the range 1-100 days, and the smallest reduced $\chi^2$ was about 6. This shows that the small timescale variations in $B_l$ seen in Fig. 4 for September 2011- January 2012 are not due to the rotation of the star, but are dominated by the errors in  $B_l$ and possible intrinsic magnetic variations.

In summary, after investigating our complete data set, we were unable to detect a rotational period shorter than 100 days, but the observed variations in both the $B_l$ and $S$-index might correspond to a rotational period of a few months.

\begin{figure}
\centering
\includegraphics[width=8 cm,angle=0] {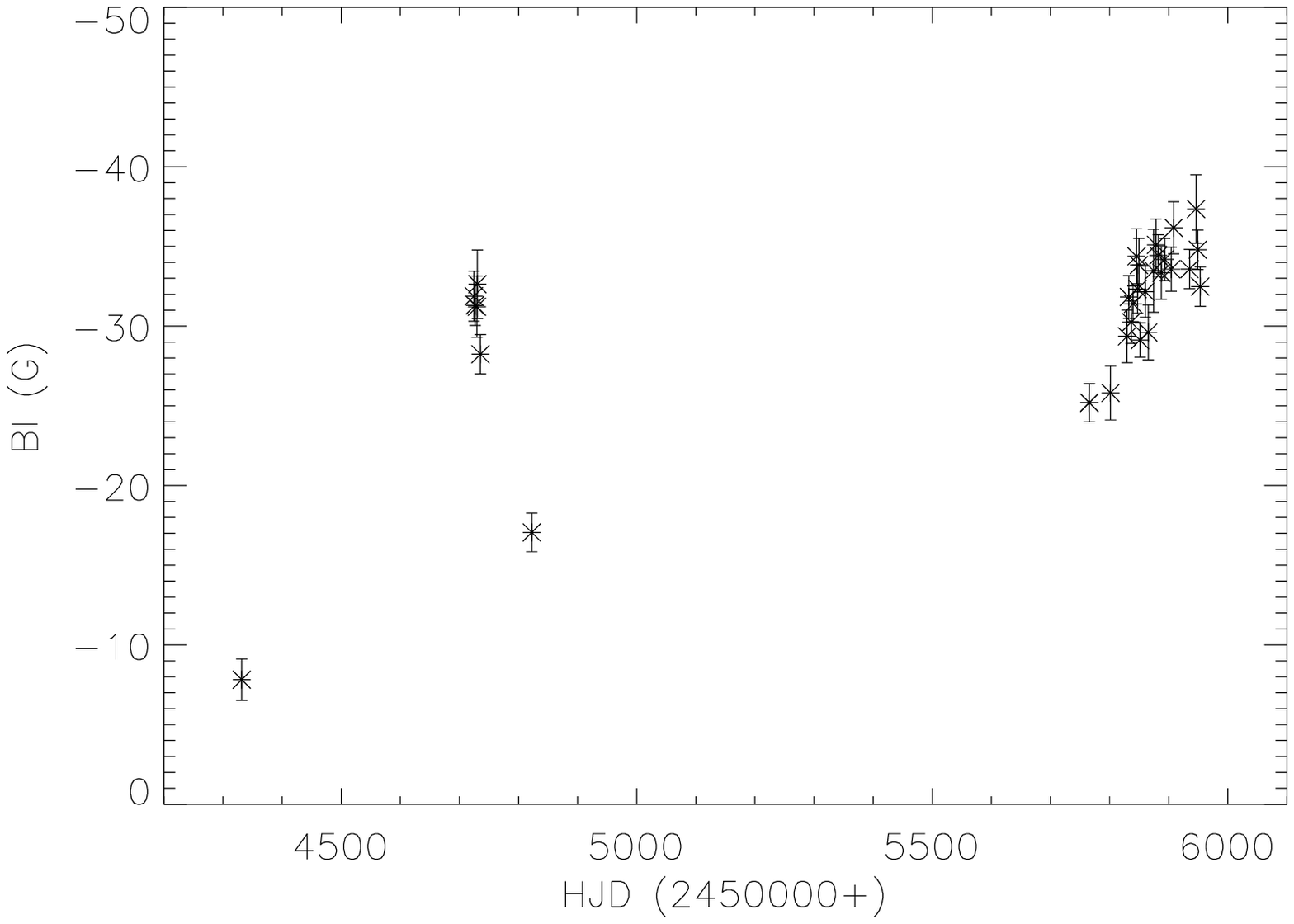}
\includegraphics[width=8 cm,angle=0] {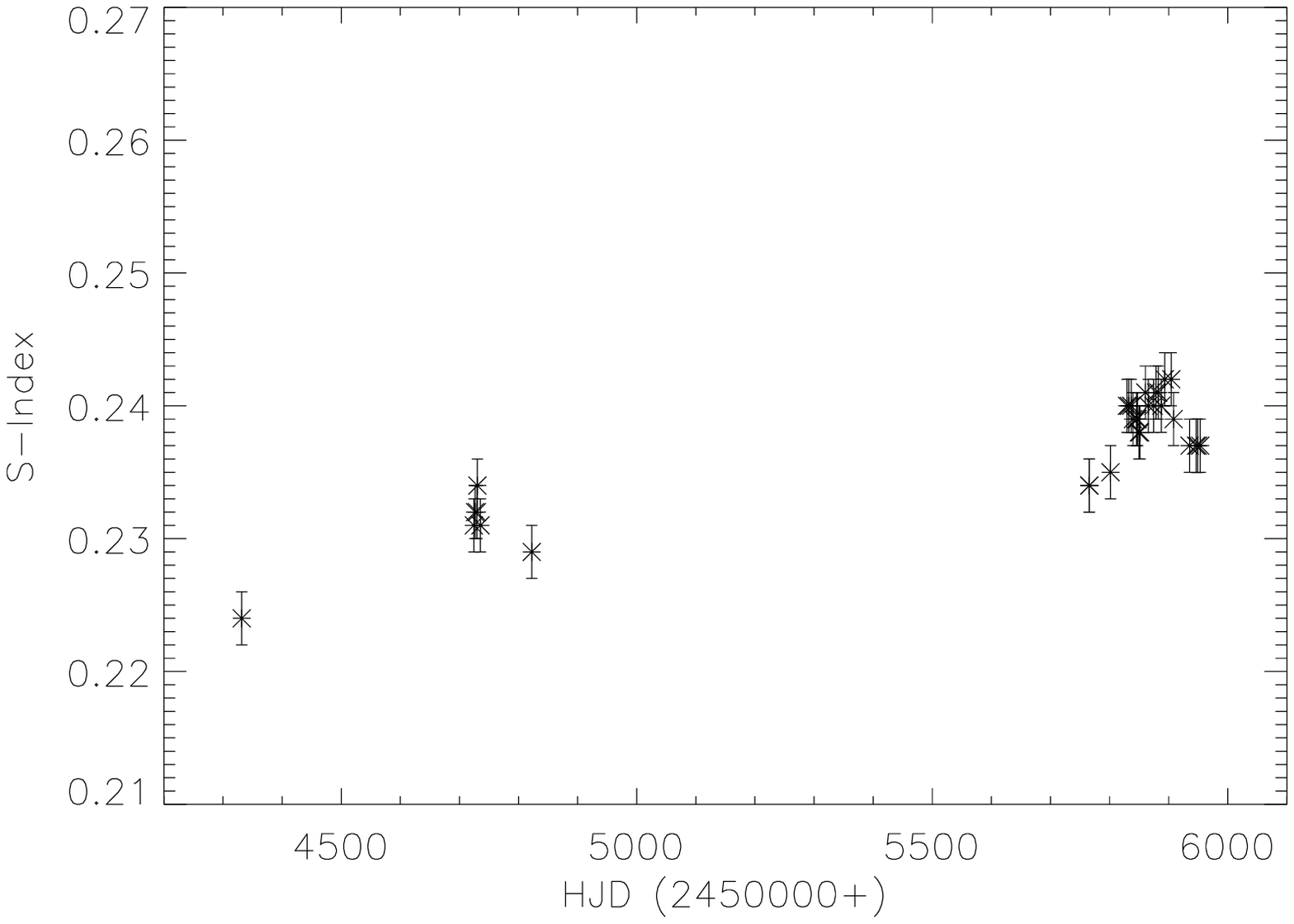}
\caption{Variations in the average longitudinal magnetic field and the $S$-index of 14 Ceti with the Julian date for the full 2007-2012 span of observations. The errors in $B_l$ are taken from Table 1, and for the $S$-index an error of 0.002 is illustrated.}
\end{figure}

\begin{figure}
\centering
\includegraphics[width=8 cm,angle=0] {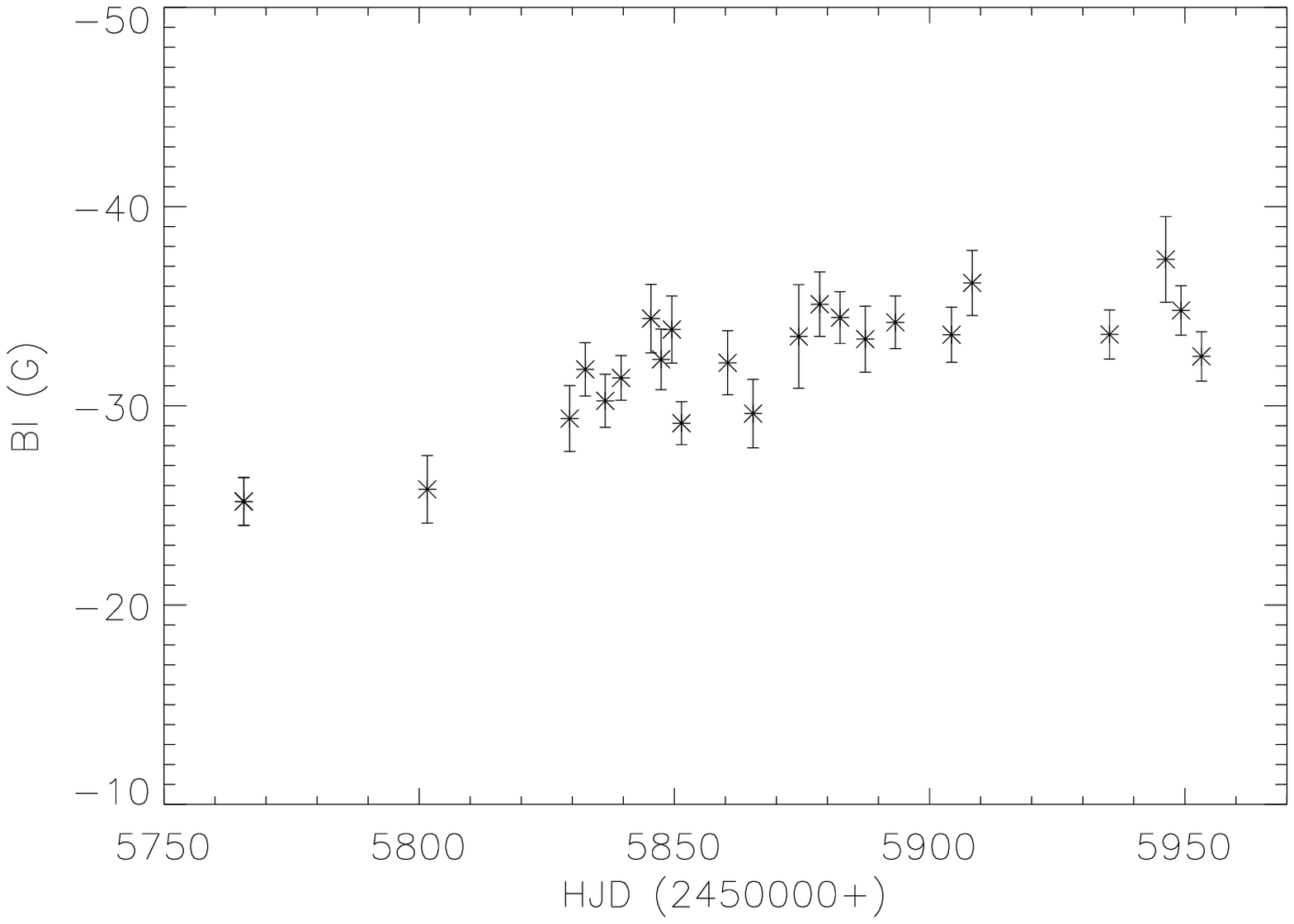}
\includegraphics[width=8 cm,angle=0] {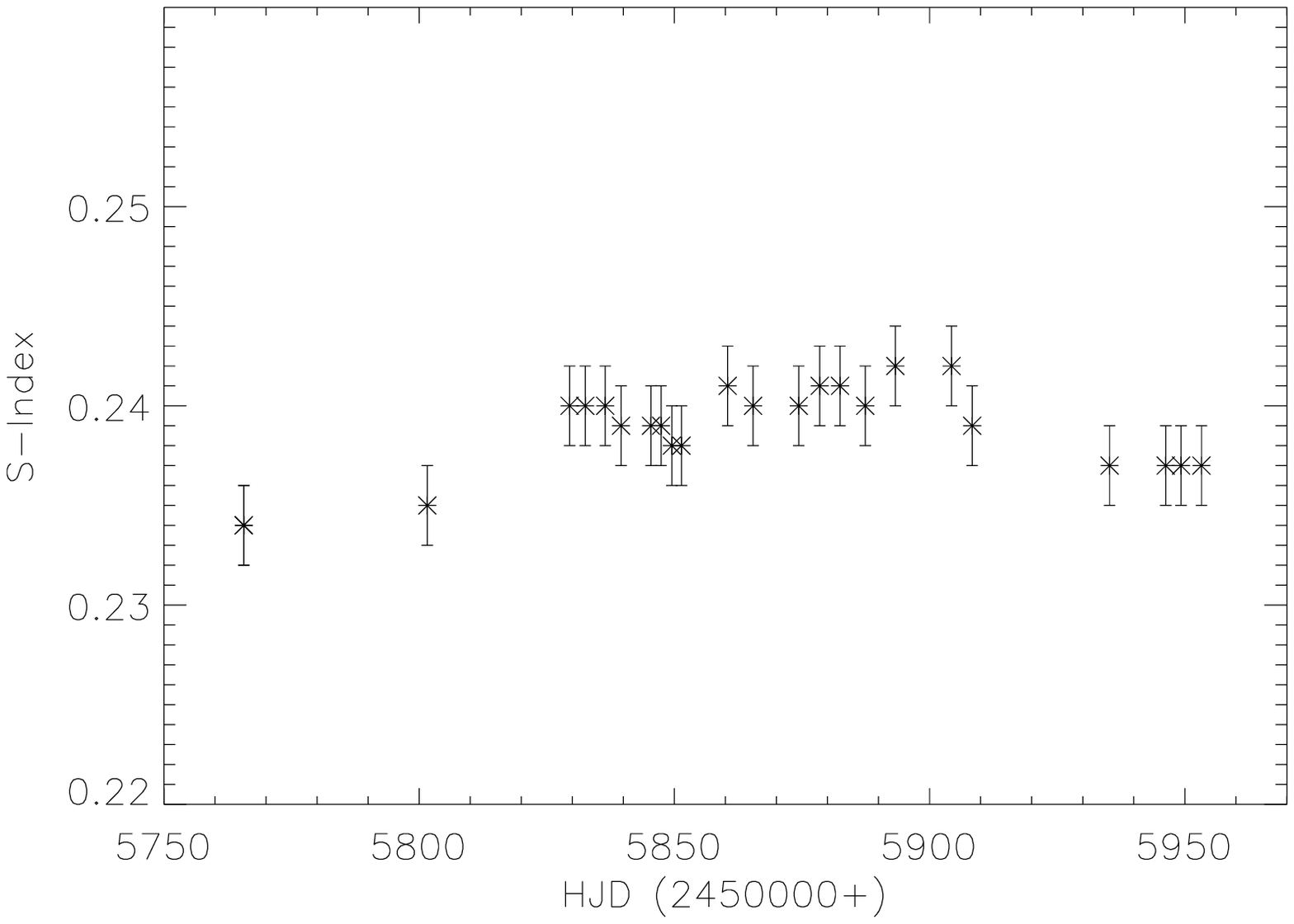}
\caption{Variations in the average longitudinal magnetic field and the $S$-index of 14 Ceti with the Julian date during the 2011-2012 season. The errors in $B_l$ are taken from Table 1, and for the $S$-index an error of 0.002 is illustrated.}
\end{figure}

\section{Abundances, fundamental parameters, and evolutionary status of 14 Ceti}

\begin{figure}
\centering
\includegraphics[width=8 cm,angle=0] {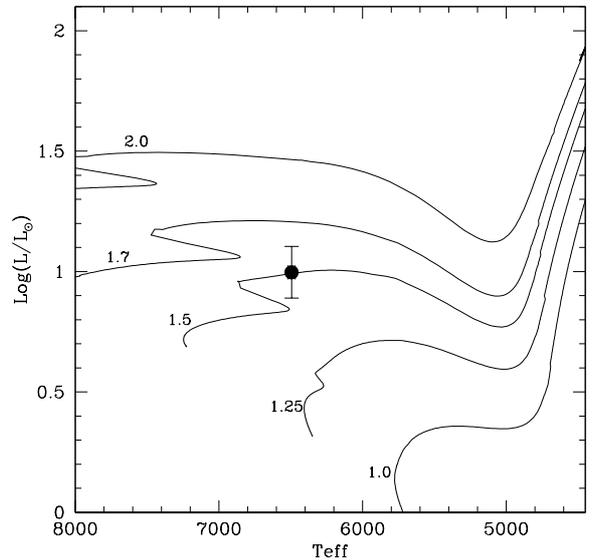} 
\caption{Position of 14 Ceti in the Hertzsprung Russell diagram, using the effective temperature derived by Van Eck et al. (in prep.) and the Hipparcos luminosity (see text). Standard evolutionary tracks at solar metallicity by Charbonnel \& Lagarde (2010) are shown from the zero-age main-sequence on, for different masses as indicated by the labels.}
\label{hrd}
\end{figure}

Van Eck et al. (in prep.) determined the stellar parameters of 14 Ceti and measured the abundances of several elements based on a detailed synthesis analysis of some of the  spectra of Narval listed in Table 1. Some results of interest are already discussed in the present paper. In particular, the derived effective temperature of $T_{\rm eff}$ = 6425 $\pm$ 25 K can be used to pinpoint the position of 14 Ceti in the Hertzsprung Russell diagram as shown in Fig. 5, where the luminosity of the star was obtained using the Hipparcos parallax of van Leeuwen (2007), the V magnitude from the 1997 Hipparcos catalogue, and the bolometric correction following Flower (1996; BC=0). When compared to the standard evolutionary tracks  of  Charbonnel \& Lagarde (2010) at solar metallicity that corresponds to the value of [Fe/H] derived from the Narval spectra, a mass of 1.5~$M_{\odot}$ was inferred for 14 Ceti. As can be seen in Fig. 5, the star is expected to have entered the Hertzsprung gap. As previously mentioned, 14 Ceti has abundances close to solar, except for Li and Be, which are depleted as previously found by L\`ebre et al. (1999) and Boesgaard et al. (2001).  As discussed in more detail in Van Eck et al. (in prep.), the Li and Be observed values can be explained by rotation-induced mixing for two different assumptions, i.e., considering that the star was either a fast rotator that has undergone no magnetic braking while on the main-sequence, or that it was initially a slow rotator that has undergone strong magnetic braking.

Since the value of $v\sin i$ is relevant to the present work, we made our own check for its determination. Feltzing et al. ( 2007) derived a macroturbulence  $v_{RT}$= 5.44 $\mathrm{\,km\,s^{-1}}$ for 14 Ceti. This value is smaller than the average
macroturbulence measured for F5 IV stars (e.g. Gray 1992). Feltzing et al. ( 2007) found that adding a rotational broadening did not improve the spectral line fitting. Adopting a macroturbulence of $v_{RT}$= 5.44 $\mathrm{\,km\,s^{-1}}$ and given the spectral resolution of Narval, we verified that a value of $v\sin i$ larger than 4 km $\mathrm{\,km\,s^{-1}}$ can be excluded.

\section{Origin of the magnetic field of 14 Ceti}

14 Ceti is the only Zeeman-detected low-mass star in the range between F0 (the coolest Ap stars, whose magnetic fields are large-scale and assumed to be of fossil origin; Johnson et al. 2006) and F7 (the hottest stars Zeeman-detected, whose magnetic fields are believed to be of dynamo origin; Donati \& Landstreet 2009).  

The detected surface magnetic field of 14 Ceti, with a longitudinal magnetic field currently reaching -30 G, is strong for an evolved star (e.g. Konstantinova-Antova 2009, Auri\`ere et al. 2009a, and 2012 in preparation). This value is one order of magnitude greater than the observed one at the surface of a few of the hottest main-sequence F7 stars that have been previously  Zeeman-detected (mean field of a few G, e.g. Donati and Landstreet, 2009), and reaches a value observed in the highly active young F7-F9 star HR 1817 (mean field of 25-50 G, Marsden 2006 and private communication). On the other hand, the magnetic strength of 14 Ceti is moderate, compared to those observed for the cool Ap stars, which are generally greater than a few hundred G (e.g. Ryabchikova et al. 2005). One has to take into account the evolution of the star, since 14 Ceti is entering the Hertzsprung gap, where the convective envelope starts deepening (which is favourable for a dynamo-driven magnetic field), and when the stellar radius increases (which would weaken a large-scale fossil magnetic field).
 
The two natural ways of explaining the magnetic strength of 14 Ceti  are then either the star is a very fast rotator with a dynamo-driven magnetic field, or it is the descendant of a cool Ap star, which is a rather slow rotator (Abt and Morrell 1985). The determination of the rotational period of 14 Ceti would have enabled us to infer directly the origin of its surface magnetic field, but we did not succeed in our investigation (Sect. 2.2), though a period longer than a few months may be present. In the following, we discuss the two different options and their implications.

\subsection{14 Ceti as a fast rotator}

At the location of 14 Ceti in the HRD, we note that the 1.5~$M_{\odot}$, $Z_{\odot}$, rotating model of Charbonnel \& Lagarde (2010), which was computed with an initial velocity of 150 $\mathrm{\,km\,s^{-1}}$ and 
presented by Van Eck et al. (in prep.), has a radius of 2.51 $R_{\odot}$, a surface velocity of 75 $\mathrm{\,km\,s^{-1}}$, and fits the observed Li and Be abundances. From these parameters and assuming solid-body rotation, we infer a rotational period of about 1.7 d.  In these conditions, the inclination should be less than 4$^\circ$ to explain the observed $v\sin i$ ($\leq$ 4 $\mathrm{\,km\,s^{-1}}$, see Sect. 3). The model also gives a convective turnover time of about four days at the bottom of the convective envelope. The inferred Rossby number at the bottom of the convective enveloppe is therefore about 0.5 and a solar-type dynamo could  be driven (Durney \& Latour 1978) at this evolutionary stage for a fast rotating star. However, at the location of 14 Ceti in the HRD, the theoretical external convection zone is thin (about 8$\%$ of the stellar radius) and its mass corresponds to that of a main-sequence star of about 1.25~$M_{\odot}$, i.e. a late FV star (using models of Charbonnel \& Lagarde, 2010). The magnetic field strength of 14 Ceti would be expected to be comparable to those that are Zeeman-detected among the most active late F main-sequence stars. 

For the hottest stars on the main-sequence where a magnetic field is believed to be dynamo-driven, a survey of solar-type stars is in progress with Narval (e. g. Petit et al. 2008). The hottest star Zeeman-detected up to now by this survey is the F7V star HD 75332, as shown in Fig. 3 of Donati \& Landstreet (2009), and the surface magnetic field is of a few G. Other late F stars have also been Zeeman studied in the case of planet-hosting stars. The first of these was $\tau$ Boo (F7V, $P_{rot}$ = 3.1 d) for which the mean magnetic field strength is measured to be a few G ( Catala et al. 2007, Donati et al. 2008, Fares et al. 2009). Several F stars observed by CoRoT also deserved deep Zeeman investigations with ESPaDOnS and Narval. 
 The active stars HD 49933 (Garcia et al. 2010, Ryabchikova et al. 2009) and WASP-12 (Fossati et al. 2010) were deeply observed but have no Zeeman detection. In the course of the present study, we also performed an observation with Narval of a kind of twin of 14 Ceti  with the same spectral class and the same X-ray emission (HD 25621, H\"unsch et al. 1998): we achieved no Zeeman detection, with a 1 $\sigma$ upper limit for $B_l$ of 2 G. 
From these surveys, only very few late F stars have been Zeeman-detected, and the magnetic field of 14 Ceti is in general one order of magnitude stronger than theirs. Only the very active F7-9 young star HR1817 (Gagn\'e et al. 1999, Budding et al. 2002) has a magnetic field that has been measured to be strong as that of 14 Ceti (HR 1817: $P_{rot} \approx $  1 d, $B_{\rm mean} \approx $ 25-50 G, Mardsden et al. 2006, 2010, Marsden 2012, private communication). $B_{\rm mean}$ is the mean unsigned magnetic field and is in general stronger than  $|B_l|^{\rm max}$ (e.g. $\xi$ Boo A, Morgenthaler et al. 2012). However, HR 1817 has the same characteristics as very active stars, including strong CaII H \& K emission cores and X-ray emission at the $10^{30}$ $\mathrm{\,erg\,s^{-1}}$ level, in contrast to 14 Ceti, which has a more quiet chromosphere and not so strong coronal activity. 
 As discussed previously, the convective envelope of 14 Ceti ($T_{\rm eff}$ = 6425 K) is relatively small and associated with a theoretical convective turnover-time of about four days (Van Eck et al., in prep.), when it is expected to be about 20 d for HR 1817 (Gagn\'e et al. 1999, $T_{\rm eff}$ = 6100 K, Marsden 2012, private communication), which is more favourable for dynamo operation.

\subsection{14 Ceti as the descendant of a cool Ap star}

Owing to its small $v\sin i$ ($\leq 4 \mathrm{\,km\,s^{-1}}$ see Sect.3), and strong $|B_l|$, 14 Ceti is a natural candidate for an Ap star descendant. Assuming solid rotation, the rotational period (in day) is longer than or equal to :
\begin{equation}
P_{\rm rot} \ge {{50.6 R_* \sin i} \over {4} },
\end{equation}
where $R_*$ is the radius of the star in $R_{\odot}$ (see e.g. Auri\`ere et al. 2007). For a representative value of 60$^\circ$ for $i$ and using the star radius inferred from the model of Charbonnel \& Lagarde (2010),
we find a rotational period longer than 27 days.

 For an Ap-star descendant, assuming conservation of magnetic flux during the post-main-sequence evolution (Stepie\'n 1993), the surface magnetic field strength is expected to decrease as the stellar radius expands as $(R_*)^{-2}$. On the main-sequence, the 1.5~$M_{\odot}$ progenitor of 14 Ceti, would be of F2 spectral type (Allen 2000) with a radius of  1.4~$R_{\odot}$. The present magnetic strength of 14 Ceti would therefore be 3.2 times weaker than that of its Ap star progenitor. If we consider that the topology of the magnetic field can be still be represented as a dipole (as in the case of the Ap-star-descendant EK Eri, which is more evolved than 14 Ceti, Auri\`ere et al. 2011), its $B_l$  variations could be fitted by an oblique rotator model (ORM, Stibbs 1950), similar to that  presented by Preston (1967) and reviewed by Auri\`ere et al. (2007). In this framework and for any geometry, the strength of the dipole is greater than
\begin{equation}
B_d \ge 3.3  B_\ell^{\rm max}.
\end{equation}
In our case, $B_d$ is greater than  120 G, and the dipole strength of the Ap star progenitor is greater than  about 370 G, i.e. stronger than the minimum magnetic field for an Ap-star of about 300 G found by Auri\`ere et al. (2007). The observed values of $B_d$ for cool Ap stars  encompass the whole strength range of magnetic fields for Ap stars (e.g.  Auri\`ere et al. 2007, studied EP UMa, an F0sp star, for which a dipole weaker than 500 G is inferred; for HD 101065, F0p, Cowley et al. 2000, derived a magnetic field strength of 2.3 kG). Since the $B_l$ of 14 Ceti does not reverse its sign, the possible ORM have limited geometries, but enable us to consider a large span of dipole strengths typical of an Ap star on the main-sequence, to explain the observed strength of about 35 G observed for $|B_l|^{\rm max}$. The search for the rotational period of 14 Ceti presented in Sect. 2.3 suggests that it could be about several months. A period $P_{rot}$  of 4 months, for example,  would imply that $v\sin i$ is smaller than 1, which is compatible with our investigation (Sect. 3).

We note that the solar abundances reported by Van Eck et al. (in prep.) for 14 Ceti, including elements that are hugely overabundant in Ap stars, might appear surprising for an Ap-star-descendant that is just leaving the main-sequence. 
However, the atmospheric-abundance anomalies induced in main-sequence Ap stars by gravitational and radiatively driven atomic diffusion are expected to be erased by the deepening convective envelope at the location of 14 Ceti in the HRD  for a  1.5$M{\odot}$, $Z{\odot}$ star (Vick \& Richard, private communication).
This was confirmed by the rotating models presented by Van Eck et al. for 
a 1.5 $M{\odot}$, $Z{\odot}$ star computed with slow initial velocity between 20 $\mathrm{\,km\,s^{-1}}$ and 50 $\mathrm{\,km\,s^{-1}}$ and assuming magnetic braking, 
as required to explain the low Li and Be abundances of 14 Ceti. In this case, the increase in both shear and meridional circulation induced by the magnetic braking  in the presence of an existing fossil magnetic field is expected to wipe out the Ap overabundances. 
\section{Conclusion}

14 Ceti is a rare star that has been Zeeman-detected in the temperature range where both fossil Ap-type and dynamo-driven magnetic fields could coexist. Evolutionary models for both of these hypotheses (corresponding to slow and fast rotators, respectively) can explain the observed abundances of 14 Ceti (Van Eck et al., in prep.). The observed strong $|B_l|$ for 14 Ceti is unexpected for a dynamo-driven magnetic field in an F-type star on the main-sequence or beyond (Sect. 4.1), and is only approached by the very active F7-F9 young star HR 1817. On the other hand, finding that $|B_l|$ is about 30 G for an Ap-star-descendant in the evolutionary state of 14 Ceti is more natural since it would correspond to the dipole strength in the range observed on the main-sequence for cool Ap stars, using an ORM of reasonable geometry and assuming conservation of the magnetic flux during the evolution. We have been unable to determine the rotational period of 14 Ceti. However, investigating our measurements obtained in 2011-2012, we could not detect any period shorter than 100 days, whereas when we consider our complete data set, the observed variations of both the $B_l$ and $S$-index suggest a rotational period of a few months.
 
If 14 Ceti is an evolved Ap star as we presently propose, its magnetic properties can be naturally explained, as in the case of EK Eri (Auri\`ere et al. 2008, 2011), by the interplay of convection with a dipolar magnetic field of fossil origin. That 14 Ceti abundances are solar and that Ap star overabundances were  wiped out on its entering the Hertzsprung gap (Van Eck et al., in prep.),  explains why no evolved Ap stars have so far been detected by mean of spectroscopy.

\begin{acknowledgements}
       We thank  the TBL team for providing service observing with Narval and the PNPS of CNRS/INSU for financial support. We acknowledge the use of the CNRS/INSU CDAB datacenter operated by the \emph{Universit\'e Paul Sabatier, Toulouse}-OMP (Tarbes, France; http://tblegacy.bagn.obs-mip.fr). The observations obtained in 2008 were funded under an OPTICON grant. R. K-A. acknowledges the possibility of working for six months in 2010 as a visiting researcher in LATT, Tarbes, under Bulgarian NSF grant DSAB 02/3/2010. She also acknowledges partial support under the Bulgarian NSF contract DO 02-85 and contract RILA 01/14. C.C. acknowledges financial support from the Swiss National Science Foundation (FNS).
\end{acknowledgements}

\end{document}